\documentclass[twocolumn,showpacs,showkeys,preprintnumbers,amsmath,amssymb]{revtex4}

\usepackage{graphicx}
\usepackage{dcolumn}
\usepackage{bm}

\begin{document}


\title{Spontaneous second order phase transition.

Amorphous branch
}

\author{Leonid S.Metlov}
\email{lsmet@fti.dn.ua}
\affiliation{Donetsk Institute of Physics and Engineering, Ukrainian
Academy of Sciences, 83114, R.Luxemburg str. 72, Donetsk, Ukraine \\
Donetsk National Univercity, 83101, Gurova avenu 14, Donetsk, Ukraine}

\thanks{}

\date{\today}

\begin{abstract}
{A version of the second order phase transition theory, in which the Nernst theorem holds automatically, 
is proposed. The theory is constructed in terms of the order parameter and the (configurational) entropy. 
It faithfully reproduces the solutions of Landau theory as well as stable existence of ordered and disordered 
states and takes into account the existence of amorphous metastable states. Finally, phenomenon of growth of 
fluctuations magnitude due to random first order transitions between stable and metastable states as their 
energies approach each other at a critical point is analyzed.}
\end{abstract}


\maketitle

The Landau theory of the second order phase transitions (PT-2) was proposed in the middle of previous 
century \cite{l37}, but interest to it still persists. The theory is widely applied for study of phase 
transitions in structural, magnetic, liquid crystal and incommensurate systems  
\cite{tscr85,tt87,t94,t12}, for study of re-orientational phase transitions \cite{bzkl76,tbknw05}. 
It had been substantially developed by modern phase fields theories \cite{akv00,esrc02,rjm09,cn12}. 
It is also used in studies of polymorph transformations of liquid – liquid type in molten silicon 
between liquid phases with differ density \cite{mwwd07}. A great many variants of PT-2 theory was 
developed in the context of amorphous material problems, for example \cite{ag65,gd58,tksw10,kmy12}. 
At the same time, it is known that this theory is not suitable to the description of amorphous 
meta-stable states \cite{kt14}. This is because amorphous states correspond to the maxima of 
thermodynamic potential, and therefore they are not absolutely stable.  To modify this theory to be 
to describe amorphous meta-stable states, we will now reconsider several foundational aspects of this 
theory.

It is well known that PT-2 theory does not satisfy the Nernst theorem. It follows from the connection 
between order parameter (OP) $\varphi$ and entropy $s$ \cite{pp79}
   \begin{equation}\label{b1}
s=-\frac{\partial f}{\partial T}=-\frac{\alpha}{2T_{c}}\varphi^2.
  \end{equation} 
where $f$ is the free energy, $T$ is the temperature,  $\alpha$ is a positive constant, 
$T_{c}$ is the critical temperature.

From here we notice that the entropy is negative for real OP ($\varphi^2>0$), that is in ordering region.  
To satisfy the Nernst theorem one can choose the free energy in the form
   \begin{equation}\label{b2}
f=f_{0}-T\frac{\alpha^2}{2bT_{c}}+\frac{1}{2}\alpha\frac{T-T_{c}}{T_{c}}\varphi^2+\frac{1}{4}b\varphi^4.
  \end{equation} 
where $b$ is another constant. 

It differs from the standard PT-2 theory only by the fact that the free term has a specific temperature 
dependence. Notice that equilibrium meaning of the OP in this case remains the same as in the reference 
theory,
  \begin{equation}\label{b3}
\varphi_{0}=0,  
\quad \varphi_{1,2}=\pm(\dfrac{\alpha}{b}\dfrac{T_{c}-T}{T_{c}})^{1/2}.
  \end{equation}

But the connection (\ref{b1}) between the OP and the entropy in this case is now different (see variant \cite{m13})
   \begin{equation}\label{b4}
s=-\frac{\partial f}{\partial T}=\frac{\alpha}{2T_{c}}(\frac{\alpha}{b}-\varphi^2),
  \end{equation}
so that the entropy becomes a positive quantity inside the ordering interval and zero at zero temperature.  

One-to-one connection (\ref{b4}) means that it is possible to choose either OP $\varphi$ or configuration 
entropy $s$, as an independent thermodynamics variable, and to express the PT-2, for example, not in terms 
of OP, but in terms of the configurational entropy. Performing the change of variables inside this interval, 
we obtain
   \begin{equation}\label{b5}
f=f_{0}^{*}-Ts+b(\frac{T_{c}}{\alpha})^2 s^2.
  \end{equation} 
All is simple and with taste. Here $f_{0}^{*}=f_{0}-\alpha^2/4b$.  

An equilibrium value of the entropy in this representation
   \begin{equation}\label{b6}
s_{1}=\frac{\alpha^2}{2bT_{c}^2}T
  \end{equation}  
after taking in account the connection (\ref{b4}) coincides by its absolute value with equilibrium values 
$\varphi_{1,2}$ (\ref{b3}) of  the classical theory. Here the coefficient $K$ plays the role of the temperature sensitivity.

Note, that the density of the internal energy in this case takes an especially simple form
   \begin{equation}\label{b7}
T=f-Ts=f_{0}^{*}+b(\frac{T_{c}}{\alpha})^2 s^2
  \end{equation}
with maximum at $s=0$, aand with no explicit temperature dependence (in equilibrium the implicit temperature 
dependence of the internal energy is via $s_{1}$). The expression for equilibrium (thermostat) temperature is 
given by the classic formula
   \begin{equation}\label{b8}
u=\frac{du}{dT}=2b(\frac{T_{c}}{\alpha})^2
  \end{equation}
which is identical to (\ref{b6}).

\begin{figure}
\includegraphics [width=3.2 in]{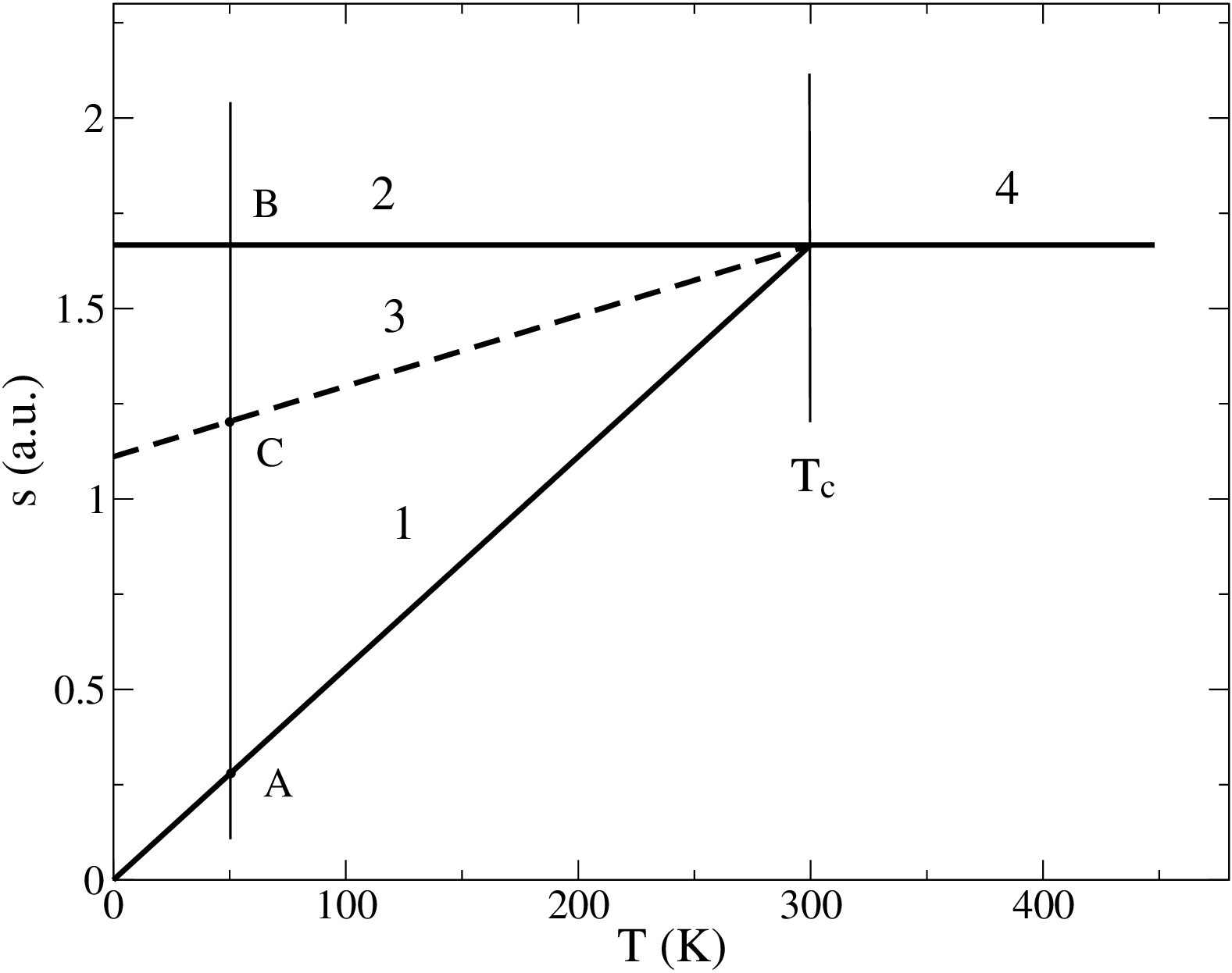}
\caption{\label{f1} . Temperature diagram of the entropy of equilibrium and metastable states: the line 1 
corresponds to the ordered state, 2 to the amorphous state, 3 to the line of maxima of free energy, 4 to 
the disordered state.}
\end{figure}

In accordance to (\ref{b6}) the equilibrium configurational entropy is zero at $T=0$, grows linearly with 
temperature and arrives to the maximal value (saturation) at the critical point $T=T_{c}$ 
(straight line 1, fig. \ref{f1}). Thus the equilibrium states $\varphi_{1,2}$  in the form (\ref{b3}) by virtue 
of quadratic connection (\ref{b4}) merge together in one state in the form (\ref{b6}), and the stable equilibrium state 
$\varphi_{0}$ in the interval $T>T_{c}$ disappears from a consideration at all.

Therefore, we propose the new free energy expression to describe the equilibrium stable states on the 
interval $T>T_{c}$
    \begin{equation}\label{b9}
f=f_{0}^{*}+\frac{\alpha^2}{2b}-T\frac{\alpha^2}{2bT_{c}}-T_{c}s+b(\frac{T_{c}}{\alpha})^2 s^2.
  \end{equation}
satisfies all the necessary requirements, namely the equilibrium entropy $s_{2}$ remains at a constant 
maximum on the interval $T>T_{c}$
    \begin{equation}\label{b10}
s_{2}=\frac{\alpha^2}{2bT_{c}}=s_{max}.
  \end{equation}  
which corresponds to a zero equilibrium in (\ref{b3}). The expression (\ref{b10}) coincides with the
expression (\ref{b6}) at the point $T = T_{c}$. Also, the temperature derivative of the free energy, 
taken with the minus sign, coincides identically with the entropy at this point.

Thus, consideration in terms of the entropy turns out somewhat more difficult (it makes it  necessary 
to determine the free energies on different temperature intervals independently). It would only be the 
question of comfort, but let us consider the reverse transition. Using the connection (\ref{b4}), let 
us express the free energy (\ref{b9}) again in terms of the order parameter
   \begin{equation}\label{b11}
f=f_{0}-T\frac{\alpha^2}{2bT_{c}}+\frac{1}{4}b\varphi^4,
  \end{equation}
which does not coincide (due to the missing quadratic term) with the general expression (\ref{b2}). 
which must be valid on the whole temperature interval. Thus, the choice of variables is not the question 
of comfort, but of the correctness of the theory, at least, in the interval $T>T_{c}$.

\begin{figure}
\includegraphics [width=3.2 in]{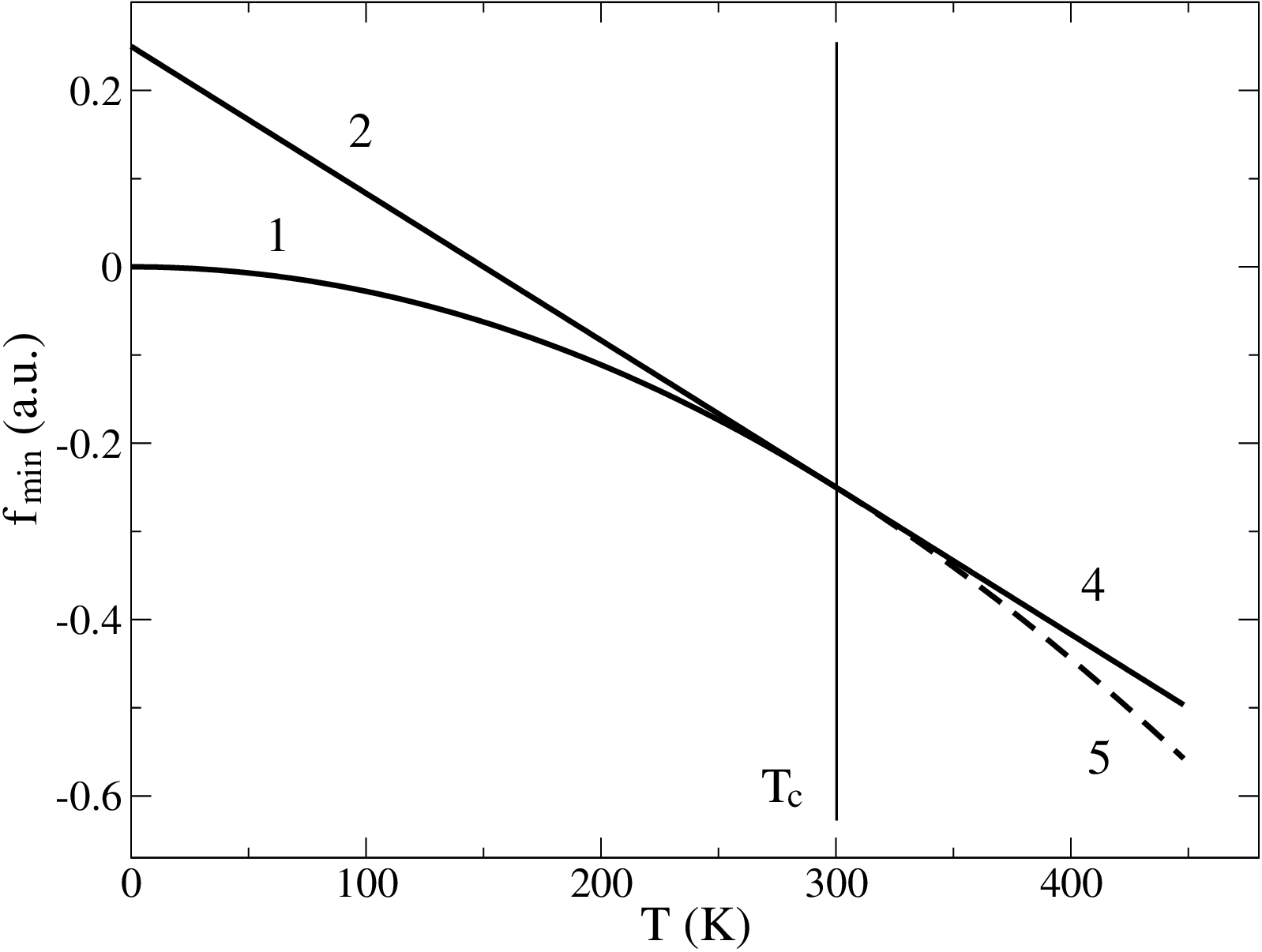}
\caption{\label{f2} Temperature dependence of the free energy at the minima: 1 corresponds to the ordered 
state, 2 to the amorphous state, 3 to the maxima of free energy, 4 to the disordered state, 5 is unphysical 
continuation of the ordered states.}
\end{figure}

Lett us show that the new form of the free energy (\ref{b9}) is physically more meaningful. At first, solution 
in the form (\ref{b9}), in the temperature interval $T>T_{c}$ can be continued to the region $T<T_{c}$. 
Indeed, there are no prohibitions on its existence in this area. But it turns out then, that in the area 
$T<T_{c}$ simultaneously exist not one (the sign of the order parameter is not taken into account), but two 
stable states, one of which is well-ordered (straight line 1, fig. \ref{f1}), and the second contains the 
frozen-in disorder or amorphous state (straight line 2, fig. \ref{f1}), which was not present in the standard 
theory. The minimum of the free energy for well-ordered states (curve 1, fig. \ref{f2}) is deeper than the 
minimum for the amorphous state (straight line 2, fig. \ref{f2}), and therefore it is the main state of the 
system. The amorphous state is meta-stable or \textquotedblleft excited\textquotedblright state.

Note that a formal extension of the free energy in the form (\ref{b5}) to the area $T>T_{c}$ in accordance with 
(\ref{b6}) leads to the unlimited entropy growth, and, consequently, in accordance to (\ref{b4}) to the imaginary 
values of OP, which is non-physical. In accordance to our concepts the system arrives at the maximum disorder 
along this degree of freedom and its further increase over this degree of freedom becomes impossible.

So, on a temperature interval below than critical point the two stable equilibrium states $s_{1}$ and $s_{2}$ 
are possible. There is a problem, how to distinguish between them, where the line of a maximum of the free energy 
passes, and what is the height of the potential barrier between these states. This issue needs additional model 
assumptions. Let us consider that the «watershed» line between the stable states (lines 1 and 2, 
fig. \ref{f1}, \ref{f2}) is a straight line (dotted line 3, fig. \ref{f1}). It goes from a critical point under 
a smaller angle, than the line for the well-ordered state (line 1, fig. \ref{f1}). We set an equation of this 
line in a form
 \begin{equation}\label{b12}
s_{3}=s_{max}^{(0)}+(s_{max}-s_{max}^{(0)})\frac{T}{T_{c}}.
  \end{equation}
Here the parameter $s_{max}^{(0)}$ determines the width of area of ordering, along the line 1 (fig. \ref{f1}).

We can restore the profile of the free energy in a power approximation from elements of the model using the 
equations for equilibrium states
 \begin{equation}\label{b13}
f=A\int(s-s_{1})(s-s_{2})(s-s_{3})ds.
  \end{equation}

\begin{figure}
\includegraphics [width=3.2 in]{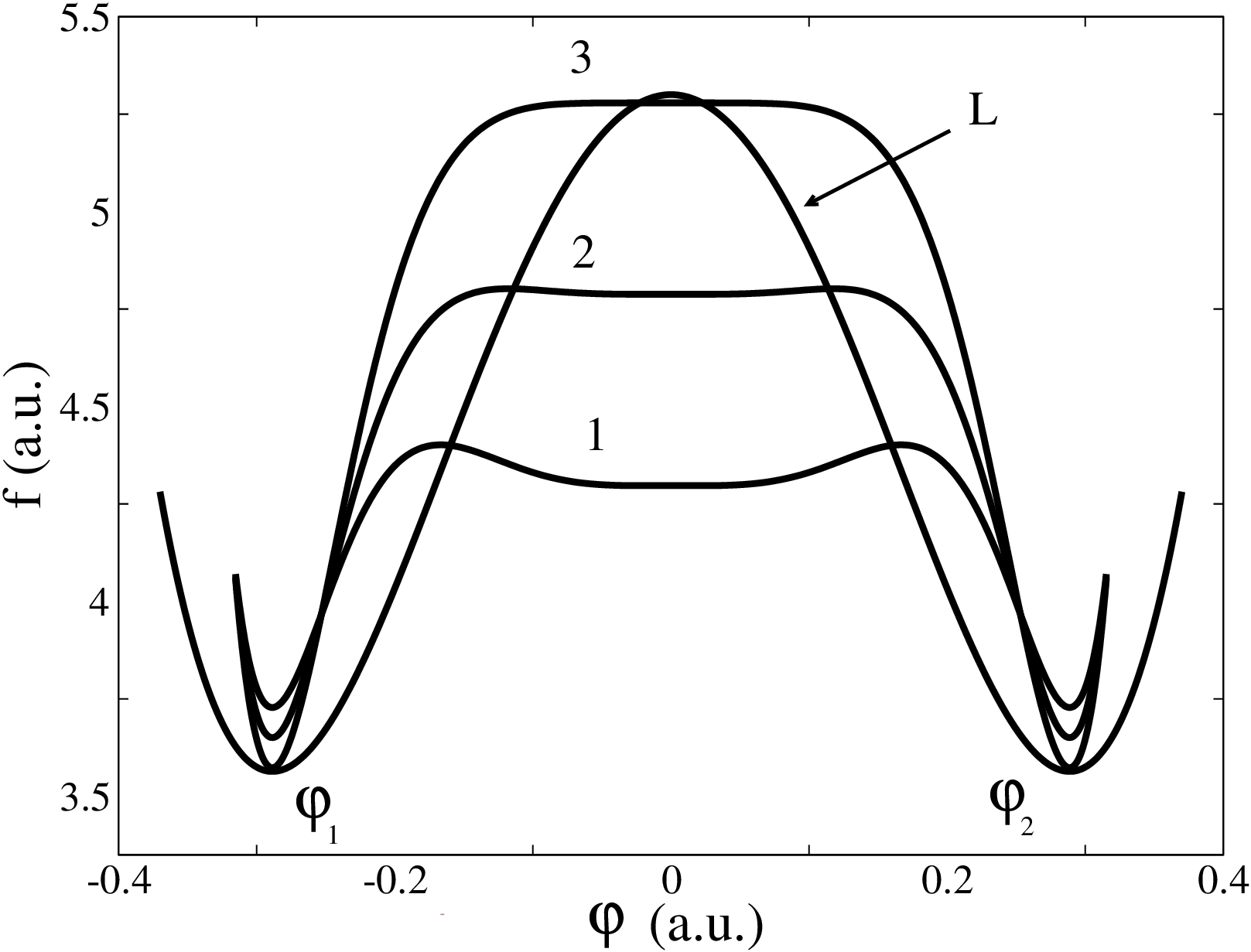}
\caption{\label{f3} The dependence of the free energy on the order parameter at the temperature $T = 50 K$. 
The curve 1 is for $s_{max 0} = s_{max} / 1.5$, the curve 2 is for $s_{max 0} = s_{max} / 1.2$, the curve 3 
is for $s_{max 0} = s_{max}$ in our free energy expression, the curve $L$ corresponds the standard theory.}
\end{figure}

The free energy curves at different values of the parameter $s_{max}^{(0)}$ are shown in fig. \ref{f3}. 
For convenience of the comparison to the standard theory, the graphics is plotted in the OP representation 
by making use of the relation (\ref{b4}). One can see that positions of equilibrium states $\varphi_{1}$ 
and $\varphi_{2}$ strictly coincide in both models. At the same time, in the region of zero OP the standard 
theory predicts unstable state of the system (maximum of the free energy), while in our case the system in 
this region is meta-stable and, in the absence of large fluctuations at low temperatures, it can exist in 
such a state for unlimited time. Notice that the minimum of the free energy in this region is quite gentle 
and the system states are in an indifferent equilibrium position in a wide region of OP values. This marks 
the possibility of realization of large number of amorphous structure variants, such as those with fractal 
landscapes of the free energy \cite{ckpuz14} that pre-determines a super-slow dynamics of system evolution 
into this region between amorphous states.

\begin{figure}
\includegraphics [width=3.2 in]{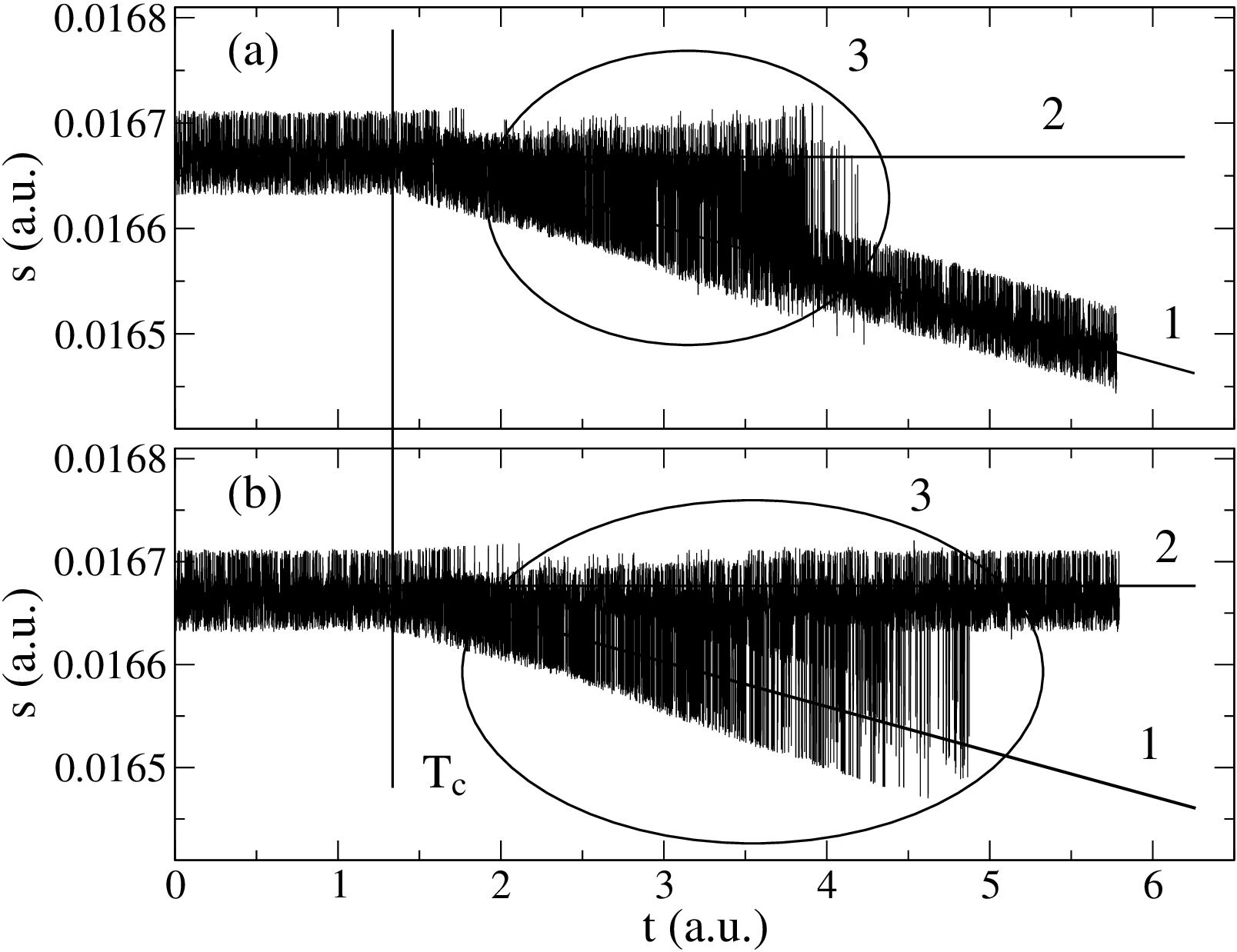}
\caption{\label{f4} Generation of long-living fluctuations at passing of critical point by the drop in the 
temperature in the case of the ordered state formation (a) and in the case of the amorphous state formation 
(b). 1 is  the line of the ordered states, 2 is the line of the amorphous states, 3 is the region of 
fluctuations growth}
\end{figure}

It is interesting to investigation the passage of the critical point by a system at heating or cooling from 
the positions of proposed theory. For example, at cooling near the critical point $T_{c}$ the system can go 
to the branch 1 (ground-state) or to the branch 2 (excited or amorphous state). Due to thermal fluctuations 
near the critical point the system will chaotically jump between these states. The probability of jumping 
into the ground-state is higher, that's why it is realized at slow cooling (fig. \ref{f4}, a). However, 
at fast cooling or at the narrow enough area of ordered states (small $s_{max}^{(0)}$) the system can remain 
in the chaotic state (quenching, fig. \ref{f4}, b).

Thus, for the ordered states  (lines 1) the chaotic states (lines 2) can be considered, as structural 
fluctuations between the chaotic states the ordered states. As fluctuations appear in the area of attraction 
of the proper nearest minimum, they will be long-living. Exactly, these long-living fluctuations determine 
the nature of growth of fluctuations near the critical point (opalescence, region 3, fig. \ref{f4}).

To simulate this process consider the evolution equation of Landau-Khalatnikov in the entropy representation
   \begin{equation}\label{b14}
\frac{\partial s}{\partial t}=-\gamma_{s}((T-T_{c})\Theta(T-T_{c})-T+2b(\frac{T_{c}}{\alpha})^2 s+Tf(s)).
  \end{equation}
where $\Theta(x)$ is theta-function, entered for comfort, $Tf(s)$ is a random function such as 
\textquotedblleft white noise\textquotedblright, which is proportional to the temperature to  simulate 
the thermal fluctuations. 

The rest of the parameters were $\alpha = 1$, $b = 0.1$, $T_{c} = 300 K$. The initial temperature of 
thermostat was taken at $T = 301 K$, which is a little higher than critical, and goes down slowly.

In the vicinity of the critical temperature, predictably, there are severe long-living fluctuations. 
Each such fluctuation is connected with penetration of the potential barrier between the main and 
the meta-stable states, that it is the first order phase transition. As long as barrier is high enough 
in the low temperature region the fluctuations are absent. When the energy levels of the main and the 
meta-stable states approach each other and the potential barrier is reduced, the magnitude of thermal 
fluctuation becomes high enough to overcome it, which gives rise to the formation of severe (structural) 
fluctuations.   This phenomenon is similar to phenomena of the random first order phase transitions in 
amorphous materials \cite{ktw89}.

It is interesting that in at the critical point itself the long-living fluctuations do not form, and the 
general level of fluctuations does not exceed the thermal background. It is related to the fat that 
distinction between the two types of steady-states is vanishing at the critical point, and they can not 
\textquotedblleft trap\textquotedblright the thermal fluctuations for each other. The growth of fluctuations, 
therefore, takes place not strictly at a critical point, but a little bit away from it to the side of the 
disordered states.

The level of fluctuations linearly increases with the distance from a critical point (due to the increase 
of the energy difference between the levels), but frequency of fluctuation diminishes. Finally, intensive 
structural fluctuations are halted, and the system gets trapped into one of the stable states. At slow 
cooling it always arrives into the main well-ordered state (branch 1, fig. \ref{f1}), at fast cooling it 
can be trapped into the meta-stable amorphous state.

Thus, in this article the second order phase transitions theory is formulated in terms of the entropy 
(configuration or structural). It reproduces, as limiting cases, all the steady-states and transitions 
of the standard theory, but also reveals new solutions, which were missing from the classic theory. It 
allows to consider the processes of super-cooling or quenching naturally. The new mechanism of fluctuations 
growth at the critical temperature is based on the closeness of the two stable states. Its important feature 
is that the maximum of fluctuations growth is not exactly at the critical point, but is displaced from it.



\begin{thebibliography}{49}
\expandafter\ifx\csname natexlab\endcsname\relax\def\natexlab#1{#1}\fi
\expandafter\ifx\csname bibnamefont\endcsname\relax
  \def\bibnamefont#1{#1}\fi
\expandafter\ifx\csname bibfnamefont\endcsname\relax
  \def\bibfnamefont#1{#1}\fi
\expandafter\ifx\csname citenamefont\endcsname\relax
  \def\citenamefont#1{#1}\fi
\expandafter\ifx\csname url\endcsname\relax
  \def\url#1{\texttt{#1}}\fi
\expandafter\ifx\csname urlprefix\endcsname\relax\def\urlprefix{URL }\fi
\providecommand{\bibinfo}[2]{#2}
\providecommand{\eprint}[2][]{\url{#2}}

\bibitem[{\citenamefont{Landau-1}(1937)\citenamefont{Landau}}]{l37}
\bibinfo{author}{\bibfnamefont{L.~D.}~\bibnamefont{Landau}},
  \bibinfo{journal}{Phys. Z. Sowjetunion} \textbf{\bibinfo{volume}{11}},
  \bibinfo{pages}{26} (\bibinfo{year}{1937}).

\bibitem[{\citenamefont{Toledano et~al.}(1985)\citenamefont{Toledano, Schmid,
  Clin, and Rivera}}]{tscr85}
\bibinfo{author}{\bibfnamefont{P.}~\bibnamefont{Toledano}},
  \bibinfo{author}{\bibfnamefont{H.}~\bibnamefont{Schmid}},
  \bibinfo{author}{\bibfnamefont{M.}~\bibnamefont{Clin}}, \bibnamefont{and}
  \bibinfo{author}{\bibfnamefont{J.~P.}~\bibnamefont{Rivera}},
  \bibinfo{journal}{Phys. Rev. B} \textbf{\bibinfo{volume}{32}},
  \bibinfo{pages}{6006} (\bibinfo{year}{1985}).

\bibitem[{\citenamefont{J.C.Toledano  and P.Toledano }(1987)}]{tt87}
\bibinfo{author}{\bibfnamefont{J.~C.}~\bibnamefont{Toledano}} \bibnamefont{and}
  \bibinfo{author}{\bibfnamefont{P.}~\bibnamefont{Toledano}},
  \emph{\bibinfo{title}{The Landau Theory of Phase Transitions: Application 
  to Structural, Incommensurate, Magnetic and Liquid Crystal Systems}} 
  (\bibinfo{publisher}{World Scientific}, \bibinfo{address}{Singapore}, \bibinfo{year}{1987}).

\bibitem[{\citenamefont{Toledano}(1994)\citenamefont{Toledano}}]{t94}
\bibinfo{author}{\bibfnamefont{P.}~\bibnamefont{Toledano}},
  \bibinfo{journal}{Ferroelectrics} \textbf{\bibinfo{volume}{162}},
  \bibinfo{pages}{257} (\bibinfo{year}{1994}).

\bibitem[{\citenamefont{Toledano}(2012)\citenamefont{Toledano}}]{t12}
\bibinfo{author}{\bibfnamefont{P.}~\bibnamefont{Toledano}},
  \bibinfo{journal}{EPJ Web of Conferences} \textbf{\bibinfo{volume}{22}},
  \bibinfo{pages}{00007} (\bibinfo{year}{2012}).

\bibitem[{\citenamefont{Belov et~al.}(1976)\citenamefont{Belov, Zvezdin,
  Kadomtseva, and Levitin}}]{bzkl76}
\bibinfo{author}{\bibfnamefont{K.~P.}~\bibnamefont{Belov}},
  \bibinfo{author}{\bibfnamefont{A.~K.}~\bibnamefont{Zvezdin}},
  \bibinfo{author}{\bibfnamefont{A.~M.}~\bibnamefont{Kadomtseva}}, \bibnamefont{and}
  \bibinfo{author}{\bibfnamefont{R.~Z.}~\bibnamefont{Levitin}},
  \bibinfo{journal}{Sov. Phys. Usp. (Sov. Adv. Phys.)} \textbf{\bibinfo{volume}{16}},
  \bibinfo{pages}{574} (\bibinfo{year}{1976}).  
  
\bibitem[{\citenamefont{Tsymbal et~al.}(2005)\citenamefont{Tsymbal, Bazalii,
  Kakazei, Nepochatykh, and Wigen}}]{tbknw05}
\bibinfo{author}{\bibfnamefont{L.~T.}~\bibnamefont{Tsymbal}},
  \bibinfo{author}{\bibfnamefont{Ya.~B.}~\bibnamefont{Bazalii}},
  \bibinfo{author}{\bibfnamefont{G.~N.}~\bibnamefont{Kakazei}},
  \bibinfo{author}{\bibfnamefont{Yu.~I.}~\bibnamefont{Nepochatykh}}, \bibnamefont{and}
  \bibinfo{author}{\bibfnamefont{P.~E.}~\bibnamefont{Wigen}},
  \bibinfo{journal}{Low Temperature Physics} \textbf{\bibinfo{volume}{31}},
  \bibinfo{pages}{227} (\bibinfo{year}{2005}).

\bibitem[{\citenamefont{Aranson et~al.}(2000)\citenamefont{Aranson,
   Kalatsky, and Vinokur}}]{akv00}
\bibinfo{author}{\bibfnamefont{I.~S.}~\bibnamefont{Aranson}},
  \bibinfo{author}{\bibfnamefont{V.~A.}~\bibnamefont{Kalatsky}}, \bibnamefont{and}
  \bibinfo{author}{\bibfnamefont{V.~M.}~\bibnamefont{Vinokur}},
  \bibinfo{journal}{Phys. Rev. Lett.} \textbf{\bibinfo{volume}{85}},
  \bibinfo{pages}{118} (\bibinfo{year}{2000}).
  
\bibitem[{\citenamefont{Eastgate et~al.}(2002)\citenamefont{Eastgate, Sethna,
  Rauscher, and Cretegny}}]{esrc02}
\bibinfo{author}{\bibfnamefont{L.~O.}~\bibnamefont{Eastgate}},
  \bibinfo{author}{\bibfnamefont{J.~P.}~\bibnamefont{Sethna}},
  \bibinfo{author}{\bibfnamefont{M.}~\bibnamefont{Rauscher}},
  \bibinfo{author}{\bibfnamefont{T.}~\bibnamefont{Cretegny}},
  \bibinfo{author}{\bibfnamefont{C.~S.}~\bibnamefont{Chen}}, \bibnamefont{and}
  \bibinfo{author}{\bibfnamefont{C.~R.}~\bibnamefont{Myers}},
  \bibinfo{journal}{Phys. Rev. E} \textbf{\bibinfo{volume}{65}},
  \bibinfo{pages}{036117} (\bibinfo{year}{2002}).

\bibitem[{\citenamefont{Rosam et~al.}(2009)\citenamefont{Rosam, Jimack, and
  Mullis}}]{rjm09}
\bibinfo{author}{\bibfnamefont{J.}~\bibnamefont{Rosam}},
  \bibinfo{author}{\bibfnamefont{P.~K.}~\bibnamefont{Jimack}}, \bibnamefont{and}
  \bibinfo{author}{\bibfnamefont{A.~M.}~\bibnamefont{Mullis}},
  \bibinfo{journal}{Phys. Rev. E} \textbf{\bibinfo{volume}{79}},
  \bibinfo{pages}{030601(R)} (\bibinfo{year}{2009}).

\bibitem[{\citenamefont{Choudhury et~al.}(2012)\citenamefont{Choudhury, and
  Nestler}}]{cn12}
\bibinfo{author}{\bibfnamefont{A.}~\bibnamefont{Choudhury}}, \bibnamefont{and}
  \bibinfo{author}{\bibfnamefont{B.}~\bibnamefont{Nestler}},
  \bibinfo{journal}{Phys. Rev. E} \textbf{\bibinfo{volume}{85}},
  \bibinfo{pages}{021602(16)} (\bibinfo{year}{2012}).

\bibitem[{\citenamefont{McMillan et~al.}(2007)\citenamefont{McMillan, Wilson,
  Wilding, and Daisenberger}}]{mwwd07}
\bibinfo{author}{\bibfnamefont{P.~F.}~\bibnamefont{McMillan}},
  \bibinfo{author}{\bibfnamefont{M.}~\bibnamefont{Wilson}},
  \bibinfo{author}{\bibfnamefont{M.~C.}~\bibnamefont{Wilding}},
  \bibinfo{author}{\bibfnamefont{D.}~\bibnamefont{Daisenberger}},
  \bibinfo{author}{\bibfnamefont{M.}~\bibnamefont{Mezouar}}, \bibnamefont{and}
  \bibinfo{author}{\bibfnamefont{G.~N.}~\bibnamefont{Greaves}},
  \bibinfo{journal}{J. Phys.: Cond. Matter} \textbf{\bibinfo{volume}{19}},
  \bibinfo{pages}{415101(41)} (\bibinfo{year}{2007}).

\bibitem[{\citenamefont{Adam et~al.}(1965)\citenamefont{Adam, and
  Gibbs}}]{ag65}
\bibinfo{author}{\bibfnamefont{A.}~\bibnamefont{Adam}} \bibnamefont{and}
  \bibinfo{author}{\bibfnamefont{J.}~\bibnamefont{Gibbs}},
  \bibinfo{journal}{J. Chem. Phys. } \textbf{\bibinfo{volume}{43}},
  \bibinfo{pages}{139} (\bibinfo{year}{1965}).

\bibitem[{\citenamefont{Gibbs et~al.}(1958)\citenamefont{Gibbs, and
  Dimarzio}}]{gd58}
\bibinfo{author}{\bibfnamefont{J.}~\bibnamefont{Gibbs}} \bibnamefont{and}
  \bibinfo{author}{\bibfnamefont{E.}~\bibnamefont{Dimarzio}},
  \bibinfo{journal}{J. Chem. Phys. } \textbf{\bibinfo{volume}{28}},
  \bibinfo{pages}{373} (\bibinfo{year}{1958}).

\bibitem[{\citenamefont{Tanaka at all}(2010)}]{tksw10}
\bibinfo{author}{\bibfnamefont{H.}~\bibnamefont{Tanaka}},
\bibinfo{author}{\bibfnamefont{T.}~\bibnamefont{Kawasaki}},
  \bibinfo{author}{\bibfnamefont{H.}~\bibnamefont{Shintani}}, \bibnamefont{and}
  \bibinfo{author}{\bibfnamefont{K.}~\bibnamefont{Watanabe}},
  \bibinfo{journal}{Nat.Mater.} \textbf{\bibinfo{volume}{9}},
  \bibinfo{pages}{324} (\bibinfo{year}{2010}).

\bibitem[{\citenamefont{Krasnuk at all}(2012)}]{kmy12}
\bibinfo{author}{\bibfnamefont{I.~B.}~\bibnamefont{Krasnuk}}, 
  \bibinfo{author}{\bibfnamefont{T.~N.}~\bibnamefont{Melnik}}, \bibnamefont{and}
  \bibinfo{author}{\bibfnamefont{V.~M.}~\bibnamefont{Yurchenko}},
  \bibinfo{journal}{Scientific records of Tavriya national university} \textbf{\bibinfo{volume}{25(64)}},
  \bibinfo{numer}{1},
  \bibinfo{pages}{193} (\bibinfo{year}{2012}).

\bibitem[{\citenamefont{Kirkpatrick et~al.}(2014)\citenamefont{Kirkpatrick,
  and Thirumalai}}]{kt14}
\bibinfo{author}{\bibfnamefont{T.~R.}~\bibnamefont{Kirkpatrick}} \bibnamefont{and}
  \bibinfo{author}{\bibfnamefont{D.}~\bibnamefont{Thirumalai}},
  \eprint{arXiv:1401.2024v1}.

\bibitem[{\citenamefont{Patashinsky and Pokrovsky}(1979)}]{pp79}
\bibinfo{author}{\bibfnamefont{A.~Z.}~\bibnamefont{Patashinsky}} \bibnamefont{and}
  \bibinfo{author}{\bibfnamefont{V.~L.}~\bibnamefont{Pokrovsky}},
  \emph{\bibinfo{title}{ Fluctuation Theory of Phase Transitions}} (\bibinfo{publisher}{Pergamon
  Press}, \bibinfo{address}{Oxford}, \bibinfo{year}{1979}).

\bibitem[{\citenamefont{Metlov}(2013)\citenamefont{Metlov}}]{m13}
\bibinfo{author}{\bibfnamefont{L.~S.}~\bibnamefont{Metlov}},
  \eprint{arXiv:1309.6791v1}.

\bibitem[{\citenamefont{Charbonneau at all}(2014)}]{ckpuz14}
\bibinfo{author}{\bibfnamefont{P.}~\bibnamefont{Charbonneau}},
\bibinfo{author}{\bibfnamefont{J.}~\bibnamefont{Kurchan}},
\bibinfo{author}{\bibfnamefont{G.}~\bibnamefont{Parisi}},
  \bibinfo{author}{\bibfnamefont{P.}~\bibnamefont{Urbani}}, \bibnamefont{and}
  \bibinfo{author}{\bibfnamefont{F.}~\bibnamefont{Zamponi}},
  \bibinfo{journal}{Nat.Commun.} \textbf{\bibinfo{volume}{5}},
  \bibinfo{pages}{3725} (\bibinfo{year}{2014}).

\bibitem[{\citenamefont{Kirkpatrick et~al.}(1989)
\citenamefont{Kirkpatrick, Thirumalai, and Wolynes}}]{ktw89}
\bibinfo{author}{\bibfnamefont{T.~R.}~\bibnamefont{Kirkpatrick}},
  \bibinfo{author}{\bibfnamefont{D.}~\bibnamefont{Thirumalai}}, \bibnamefont{and}
  \bibinfo{author}{\bibfnamefont{P.~G.}~\bibnamefont{Wolynes}},
  \bibinfo{journal}{Phys. Rev. A} \textbf{\bibinfo{volume}{40}},
  \bibinfo{pages}{1045} (\bibinfo{year}{1989}).

\end{thebibliography}

\end{document}